\documentclass[twocolumn,           % Format : preprint, twocolumn
               showpacs,            % Pacs : showpacs, noshowpacs
               preprintnumbers,     % Preprint: preprintnumbers,
               aps,                 % Society: ...
               prd,                 % Journal Style : pra, prb, prc, prd, pre,
               a4paper,             % Size : a4paper, ...
               superscriptaddress,  % Affiliation (Title) : groupedaddress,
               nofootinbib,         % Footnote: footinbib, nofootinbib
               tightenlines,        % Remove additional spaces in a line
               floats,floatfix      % Floating pictures and tables
               ]{revtex4}

\usepackage{graphicx}
\usepackage{latexsym}
\usepackage{amsmath,amssymb}        % amssymb includes amsfonts
\usepackage[draft=false]{hyperref}

\begin{document}

%%--- DRAFTCOPY --------------------------------
%% Prints a large "DRAFT" diagonally across each page
%% Does not show up in TeXview
%% \typeout{Prints "DRAFT" on each page; does not show in TeXView}
%% \special{!userdict begin /bop-hook{gsave 200 30 translate
%% 65 rotate /Times-Roman findfont 216 scalefont setfont
%% 0 0 moveto 0.90 setgray (DRAFT) show grestore}def end}
%%------------------------------------------------

%======================================%
%<<<<<<<<<<<< TITLE PAGE >>>>>>>>>>>>>>%
%======================================%

\title{The WMAP normalization of inflationary cosmologies} 
\author{Andrew R.~Liddle} 
\affiliation{Astronomy Centre, University of Sussex, Brighton BN1 9QH, 
United Kingdom}  
\author{David Parkinson} 
\affiliation{Astronomy Centre, University of Sussex, Brighton BN1 9QH, 
United Kingdom} 
\author{Samuel M.~Leach}
\affiliation{SISSA--ISAS, Astrophysics Sector, Via Beirut 4,
34014 Trieste, Italy}
\author{Pia Mukherjee} 
\affiliation{Astronomy Centre, University of Sussex, Brighton BN1 9QH, 
United Kingdom} 
\date{\today} 
\pacs{98.80.-k \hfill astro-ph/0607275} 
\preprint{astro-ph/0607275} 
 
\begin{abstract} 
We use the three-year WMAP observations to determine the normalization
of the matter power spectrum in inflationary cosmologies. In this
context, the quantity of interest is not the normalization
marginalized over all parameters, but rather the normalization as a
function of the inflationary parameters $n$ and $r$ with
marginalization over the remaining cosmological parameters. We compute
this normalization and provide an accurate fitting function. The
statistical uncertainty in the normalization is 3 percent, roughly
half that achieved by COBE. We use the $k$--$\ell$ relation for the
standard cosmological model to identify the pivot scale
for the WMAP normalization. We also quote the inflationary energy
scale corresponding to the WMAP normalization.
\end{abstract} 
 
\maketitle

\section{Introduction} 
 
One of the legacy products of the COBE satellite \cite{cobe} was a
determination of the normalization of the matter power spectrum, which
to this day is often referred to as the COBE normalization.
Surprisingly, only limited effort has been made so far to extract the
equivalent result from Wilkinson Microwave Anisotropy Probe (WMAP)
observations \cite{wmap3,wmap3cal}, with the papers to date quoting
only the normalization marginalized over all cosmological
parameters. For inflationary cosmologies this is not quite the
quantity required; inflation models predict the spectral index and
tensor-to-scalar ratio, and hence should be normalized assuming these
parameters are fixed. Because the normalization does depend
significantly on those parameters even across just the region allowed
by the data, marginalization over those parameters significantly
worsens the apparent uncertainty, as well as losing the information on
how the normalization correlates to these parameters.

The COBE normalization of inflationary cosmologies was given, from the
four-year data, by Bunn, Liddle and White \cite{BLW}. They used a
data-analysis methodology described by Bunn and White \cite{BW}. Their
treatment considered both the then-popular critical-density cosmology,
and also the flat dark-energy dominated cosmologies favoured
today. The results were somewhat confusing, as they used a definition
of the normalization that depended on the value of $\Omega_0$ (see
e.g.~Ref.~\cite{MRS}), and also quoted the present-day normalization
from which a growth factor correction would have to be subtracted to
generate the primordial amplitude predicted by inflation.

Nevertheless, the result had a statistical uncertainty of only 7
percent, and has been frequently used to constrain inflationary
models. In single-field models it simply fixes the overall
normalization of the potential; in a one-parameter case such as $V =
\frac{1}{2} m^2 \phi^2$ this completely fixes the model. In hybrid
inflation models the constraint may be less trivial, as the
instability point ending inflation may not scale directly with the
overall normalization. In models such as the curvaton model \cite{LW},
where adiabatic perturbations are generated after inflation from
isocurvature ones, the normalization constrains a combination of model
parameters including the initial value of the curvaton field.

No inflation theory purports to predict the perturbation amplitude at
anything like the 7 percent level already achieved by
COBE. Nevertheless, we argue that the WMAP normalization is an
important legacy result of the WMAP survey.  Further, it is bizarre
that the currently quoted normalization uncertainties from WMAP hardly
improve on the quoted COBE level at all, due to the marginalization
over inflationary parameters.  It is also an opportunity to quote the
result in more modern model-independent terminology.

\section{Normalization calculation}

The normalization calculation is straightforward. The inflationary
parameters are $n$ and $r$, giving the density perturbation spectral
index and the tensor-to-scalar ratio in standard conventions; a
typical inflation model would predict values for those and hence
should be normalized assuming those fixed values. We carried out a
series of runs of the CosmoMC code \cite{Lewis:2002} in which $n$ and
$r$ are fixed at different values across a grid, making 28 evaluations
combining $n=0.90, 0.925, 0.95, 0.975, 1, 1.025, 1.05$ and $r=0, 0.25,
0.5, 0.75$. This comfortably covers the values of these parameters
favoured by current observations from WMAP combined with other
data. In our analysis we consider WMAP data alone, including both
temperature and polarization data. A power-law primordial spectrum is
assumed throughout and the tensor spectral index fixed by the
single-field inflationary consistency equation (see
e.g.~Ref.~\cite{LL}). The parameters allowed to vary and subsequently
marginalized over are the matter density $\Omega_{\rm m}$, the baryon
density $\Omega_{\rm b}$, the Hubble constant $h$, and the optical
depth $\tau$.

The normalization parameter, denoted $A_{\rm s}$ in CosmoMC, is
equivalent to ${\cal P}_{\cal R}$ of the textbook by Liddle and Lyth
\cite{LL}, evaluated at the comoving scale $0.05 \, \mbox{Mpc}^{-1}$
at some early epoch while the perturbations are still
superhorizon. This quantity gives the normalization in a way which is
independent of the ultimate value of $\Omega_0$ at the present. We use
the related quantity $\delta_{{\rm H}} \equiv 2{\cal P}_{\cal
R}^{1/2}/5$.

The simplest way to define our terminology is to relate it to the
perturbations from slow-roll inflation models. There $\delta_{{\rm
H}}$ is given in terms of the Hubble parameter $H$ and scalar field
velocity $\dot{\phi}$ by (see e.g.~Ref.~\cite{LL})
\begin{equation}
\delta_{{\rm H}}^{{\rm inflation}} \simeq \frac{H^2}{5\pi
\left|\dot{\phi}\right|} \,.
\end{equation}
The right-hand side is to be evaluated at the time when the
normalization scale crossed outside the horizon during inflation,
which can be estimated following Ref.~\cite{LiLe}. The spectral index
and tensor-to-scalar ratio are given in the slow-roll approximation by
\begin{eqnarray}
n - 1 & \simeq & -6\epsilon + 2 \eta \; \\
r & \simeq & 16\epsilon \,,
\end{eqnarray}
where the slow-roll parameters are defined from the potential
$V(\phi)$ by
\begin{equation}
\epsilon = \frac{m_{{\rm Pl}}^2}{16\pi} \, \left( \frac{dV/d\phi}{V}
\right)^2 \quad ; \quad \eta = \frac{m_{{\rm Pl}}^2}{8\pi} \,
\frac{d^2V/d\phi^2}{V} \,.
\end{equation}

To compare with the COBE normalization of Ref.~\cite{BLW}, one should
simply look at the critical-density result given in that case, 
\begin{equation}
\label{e:cobe}
\delta_{\rm H}^{\rm COBE} = 1.91 \times 10^{-5}
\, \frac{\exp\left[1.01(1-n)\right]}{\sqrt{1+0.75\hat{r}}} \,.
\end{equation}
Here $\delta_{{\rm H}}$ is quoted at the present horizon scale $k =
aH$, and we have used the notation $\hat{r}$ to indicate use of an
outdated definition of the tensor-to-scalar ratio, connected to the
modern one by $r = 1.29 \hat{r}$.  That the dependence on $n$ is well
approximated by an exponential implies that there is a scale where the
normalization becomes independent of $n$, the pivot scale, here given
by $k_{{\rm pivot}}^{{\rm COBE}} = e^{2.02} aH$.

\begin{figure}[t]
\includegraphics[width=0.9 \linewidth]{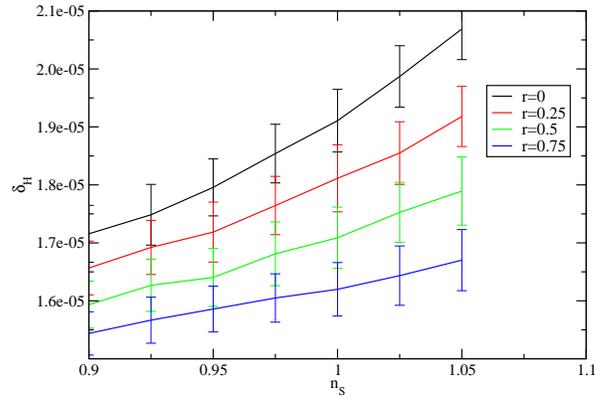}
\caption{The WMAP3 normalization $\delta_{{\rm H}}$ as a function of
$n$ for the four choices of $r$, specified at the scale $0.05 \, {\rm
Mpc}^{-1}$. The uncertainties shown are the standard deviations
inferred from the Markov chains.}
\label{f:norm}
\end{figure}

Our result for the normalization is shown in
Fig.~\ref{f:norm}. Generally we find excellent agreement with the COBE 
normalization; in particular for $n=1$ and $r=0$ we get $\delta_{{\rm
H}} = 1.91 \times 10^{-5}$, matching the value from the COBE fitting
function. Our findings are thus consistent with the conclusion of
Ref.~\cite{wmap1bennett} that the WMAP measurements of the large
angular scale anisotropy show no systematic discrepancy from the COBE
measurements.

We find that the results are well fit by the function
\begin{equation}
\label{e:wmap}
\delta_{{\rm H}}^{{\rm WMAP}} = 1.927 \times 10^{-5} \frac{\exp
\left[(-1.24+ 1.04 r) (1-n) \right]}{\sqrt{1+0.53r}} \,,
\end{equation}
which is the main result of this paper. Here $\delta_{{\rm H}}^{{\rm
WMAP}}$ is specified at $k=0.05\, {\rm Mpc}^{-1}$ and hence the
coefficients cannot be directly compared with those in
Eq.~(\ref{e:cobe}). The criteria for obtaining the fit parameters was
the minimization of the maximum relative error over the $n$ and $r$
values studied.  Similar results are obtained by minimizing the
average deviation or the chi-squared.  The fitting function is
accurate to within one percent for all the values we considered.

The form of the fitting function in Eq.~(\ref{e:wmap}) is motivated by
the same criteria as for COBE. If there is a pivot scale where the
normalization becomes independent of $n$, the dependence on $n$ should
be exponential to reflect that translation of scales. The introduction
of $r$ suppresses the matter power spectrum, the coefficient of order
unity required because $r$ is defined by the relative tensor and
scalar power spectra rather than precisely by their relative effect on
the CMB.

However, with WMAP it turns out that there is not a unique pivot scale
independent of $r$; as $r$ is introduced the curves shown in
Fig.~\ref{f:norm} become noticeably flatter. The pivot scale when
$r=0$ is $k_{{\rm pivot}}^{{\rm WMAP}, r=0} = e^{-2.48} \times 0.05 \,
{\rm Mpc}^{-1} = 0.004 \, {\rm Mpc}^{-1}$, increasing to $0.02 \, {\rm
Mpc}^{-1}$ when $r = 0.75$ (when $r$ is large, the low multipoles
contain less information about the matter power spectrum
normalization). The $r$-dependence in the exponential is necessary to
allow for this in the fitting function.

The statistical uncertainty on $\delta_{{\rm H}}^{{\rm WMAP}}$ is
independent of $n$ and $r$ to a good approximation and is 3 percent
(at 68 percent confidence), corresponding to about twice the accuracy
of COBE. The two results are entirely consistent with one
another. This uncertainty is also about half of that obtained when
marginalizing the amplitude over the posterior distribution of $n$ and
$r$; the WMAP team quote $\delta_{{\rm H}} = (1.83 \pm 0.10) \times
10^{-5}$ for the fully-marginalized amplitude in inflationary models
\cite{lambda} (this number corresponding to the scale $0.002 \, {\rm
Mpc}^{-1}$).

The precision of the normalization is also affected by the overall
calibration uncertainty of the WMAP data. This is estimated to be 0.5
percent in the WMAP3 data \cite{wmap3cal}, the same level reported for
the first-year data. That number was obtained \cite{wmap1hin} by
applying an iterative algorithm that simultaneously fits for
calibration parameters and sky maps; using detailed simulations of the
time-ordered data they arrive at a conservative estimate of 0.5
percent for the absolute calibration uncertainty. This calibration
uncertainty is well below the statistical uncertainty reported above
(0.5 percent is the appropriate number for $\delta_{\rm H}$, and would
correspond to a 1 percent uncertainty in the power spectrum).

\begin{figure}[t]
\includegraphics[width=0.9 \linewidth]{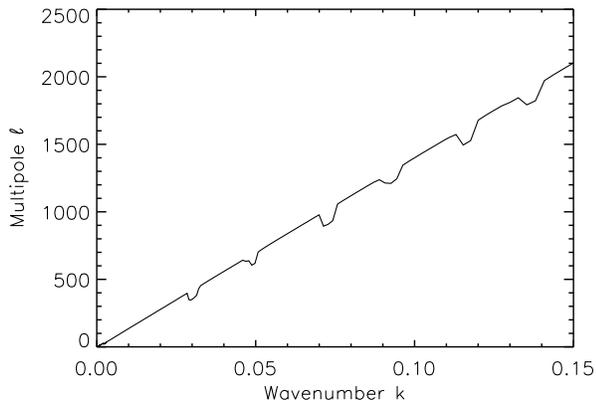}
\caption{The relation between multipole $\ell$ and wavenumber $k$ for
the best-fit cosmology to WMAP3 data \cite{wmap3}.}
\label{f:ktol}
\end{figure}

Having evaluated the effective pivot scale (for different $r$ values),
it is interesting to know what multipole value it corresponds to, in
order to see which part of the data drives the result. To make this
relation, we ran a series of CMB spectrum computations using spike
power spectra ($\Delta \ln k = 0.02$) at different $k$, using the
best-fit WMAP cosmological parameters $\{\Omega_{\rm b}h^2,\Omega_{\rm
m}h^2, h,\tau \}=\{0.0223,0.1263,0.73,0.088 \}$.  Such a sharp input
spectrum gets smeared by projection effects, but we can still pick off
the corresponding $\ell$ value where the transfer of power peaks.  (If
we instead take the mean $\ell$, rather than the peak location, the
$\ell$ value corresponding to a given $k$ is reduced by about 15\%.)

The correspondence is shown in Fig.~\ref{f:ktol}; ignoring the
glitches which coincide with the CMB temperature acoustic troughs, the
relation is well fit by
\begin{equation}
\label{e:kell}
\ell = 14,000 \; \frac{k}{{\rm Mpc}^{-1}}, \,\, \ell\gg 1.
\end{equation}
Using this, we find that the pivot scale for $r=0$ corresponds to
$\ell \simeq 60$, increasing to about 280 for the largest $r$ values
considered. We conclude that even with WMAP the majority of the
constraining power on the amplitude is at fairly low $\ell$, which is
why the improvement with respect to COBE is limited. From a physical
point of view, cosmic variance limited measurements of the
Sachs--Wolfe effect, because of its microphysics independence, carry
the most statistical weight in this context.

On the other hand, the pivot scale for the well-determined combination
$\delta_{{\rm H}}e^{-\tau}$ is found to be $0.033 \, {\rm Mpc}^{-1}$
for $r=0$, corresponding to an $\ell$ of about 460, more or less in
the middle of the WMAP range. Therein lies the justification for
currently using $0.05 \, {\rm Mpc}^{-1}$ for the scalar pivot scale.
Currently the CMB acoustic peak region, where the overall level of
anisotropy can be best determined, suffers from a perfect degeneracy
between the amplitude and the optical depth, broken only towards low
$\ell$. A better measurement of $\tau$ from CMB polarization will help
break this degeneracy, allowing a better determination of
$\delta_{{\rm H}}$ and moving its pivot scale up to higher $\ell$.
$\delta_{{\rm H}}e^{-\tau}$ is currently known to an accuracy of about
$1.3$ percent, so that external constraints on $\tau$ could lead to at
most a factor two improvement in our knowledge of $\delta_{{\rm H}}$
with present CMB data.

In single-field slow-roll inflation models, the perturbation
normalization directly gives the inflationary energy scale in terms of
the slow-roll parameter $\epsilon$ or, equivalently, the
tensor-to-scalar ratio. Following the usual calculation (see
e.g.~Refs.~\cite{BLW,LL}), we find that the energy scale as a given
mode crossed the horizon was
\begin{equation}
\frac{V^{1/4}}{M_{{\rm Pl}}} = \left( \frac{75\pi^2}{8} \right)^{1/4}
r^{1/4} \, \delta_{{\rm H}}^{1/2} \,,
\end{equation}
where $M_{{\rm Pl}} = 2.436 \times 10^{18} \, {\rm GeV}$ is the
reduced Planck mass. Substituting in the WMAP normalization of
Eq.~(\ref{e:wmap}) gives
\begin{eqnarray}
\frac{V_*^{1/4}}{1 \, {\rm GeV}}& =& (3.32 \pm 0.05) \times 10^{16}
\times r^{1/4} \times \\ && \quad \frac{\exp \left[(-0.62+ 0.52 r)
(1-n) \right]}{(1+0.53r)^{1/4}} \,, \nonumber
\end{eqnarray}
where `*' indicates that this corresponds to when the scale $0.05 \,
{\rm Mpc}^{-1}$ crossed outside the horizon during inflation. This
agrees with previous results, with a smaller uncertainty in the
prefactor. Nevertheless, one cannot expect the other terms to be
measured with anything like the precision of the prefactor, the
uncertainty on the $r^{1/4}$ being likely to be the limiting term.  In
particular applications one should also check whether slow-roll
corrections are comparable to or greater than the statistical
uncertainty.

\section{Conclusions}

We have determined the density power spectrum normalization from WMAP,
as a function of the inflationary parameters $n$ and $r$. This WMAP
normalization is the one appropriate to specific inflation models, and
its statistical uncertainty improves on the COBE normalization by a
factor of approximately two. Explicit examples of how to use the
normalization to constrain inflation models are given in
Refs.~\cite{BLW,LL}.

\begin{acknowledgments}

A.R.L., P.M.\ and D.P.\ were supported by PPARC. S.M.L.\ acknowledges
a visit to Sussex supported by PPARC. We acknowledge use of the UK
National Cosmology Supercomputer (COSMOS) funded by Silicon Graphics,
Intel, HEFCE and PPARC, and of the Legacy Archive for Microwave
Background Data Analysis (LAMBDA). Support for LAMBDA is provided by
the NASA Office of Space Science.
\end{acknowledgments}

%======================================% 
%<<<<<<<<<<<< BIBLIOGRAPHY >>>>>>>>>>>>% 
%======================================% 

%%%%%%%%%%%%%%%%%%%%%%%%%%%%%%%%%%%%%%%%%%%%%%%%%
\end{document}